\documentclass[12pt]{article}

\usepackage{epsfig}

\def\lr{\left( }
\def\rr{\right) }
\def\bea{\begin{eqnarray}}
\def\eea{\end{eqnarray}}

\def  \bcen     {\begin{center}}
\def  \ecen     {\end{center}}
\def  \beq      {\begin{equation}}
\def  \eeq      {\end{equation}}
\def  \beqa     {\begin{eqnarray}}
\def  \eeqa     {\end{eqnarray}}

\def  \bfleft   {\begin{flushleft} }
\def  \efleft   {\end{flushleft} }
\def  \bfright  {\begin{flushright} }
\def  \efright  {\end{flushright} }

\def  \et       {\ \not\!\!\! E_T}

\def  \pptoqqbjjll  {p p \to Q_H Q'_H \to j j \ell^+ \ell^- \not\!\!
  E_T} 

\def  \signatureone  {\ell^\pm \ell^\mp j j \not\!\! E_T}
\def  \signaturetwo  {\ell^\pm \ell^\pm j j \not\!\! E_T}
\begin{document}

\def\thesection {\Roman{section}}
\def\thesubsection {\Alph{subsection}}

\renewcommand{\thefootnote}{\fnsymbol{footnote}}

\begin{flushright}
LPSC 09-167 \\
LYCEN 2009-12 
%arXiv:yymm.nnnn [hep-ph]
\end{flushright}
\vskip 2cm
\bcen
%%%%%  Title =========================%%
{\Large \bf \boldmath 
Dileptonic signatures of $T$-odd quarks at the LHC} 
\vskip 2cm
%%%%%% Authors =======================%%
{\bf 
Giacomo Cacciapaglia$^1$\footnote{g.cacciapagalia@ipnl.in2p3.fr},
S. Rai Choudhury$^2$\footnote{srai.choudhury@gmail.com},
Aldo\ Deandrea$^1$\footnote{deandrea@ipnl.in2p3.fr},  \\ \vspace{1ex} 
Naveen\ Gaur$^3$\footnote{gaur.nav@gmail.com}, 
Michael\ Klasen$^4$\footnote{klasen@lpsc.in2p3.fr}
}
%%%%%% Affiliations =======================%%
\vskip 1.2cm
{\sl 
$^1$ Universit\'e de Lyon, F-69622 Lyon, France; Universit\'e Lyon 1,\\
CNRS/IN2P3, UMR5822 IPNL, F-69622 Villeurbanne Cedex, France \\ 
$^2$ Center for Theoretical Physics (CTP), Jamia Millia University,
Delhi \& 
Indian Institute of Science Education \& Research (IISER),
Govindpura, Bhopal, India \\  
$^3$Department of Physics \& Astrophysics, University of Delhi, Delhi -
110007, India \\
$^4$Laboratoire de Physique Subatomique et de Cosmologie,
Universit\'e Joseph Fourier/CNRS-IN2P3/INPG, 53 Avenue des Martyrs,
38026 Grenoble, France \\
}
\ecen
\vskip 1cm
%+++++++++++++++++++++++++++++++++++++++++++++++++++++++++++
% Abstract
\thispagestyle{empty}
\begin{abstract}
Little Higgs models are often endowed with a $T$-parity in order
to satisfy electroweak precision tests and give at the same time a
stable particle which is a candidate for cold dark matter. This type
of models predicts a set of new $T$-odd fermions in addition to the
heavy gauge bosons of the Little Higgs models, which may show
interesting signatures at colliders. In this paper, we study the
signatures of strong and electroweak pair production of the first two generations of $T$-odd quarks
at the LHC. We focus on the dileptonic signatures 
 (a) $p p \to \ell^\pm \ell^\mp j j \et$ (opposite-sign dileptons)
 and (b) $p p \to \ell^\pm \ell^\pm j j \et$ (same-sign dileptons).  
\end{abstract}
%+++++++++++++++++++++++++++++++++++++++++++++++++++++++++++
\vfill \eject 

\renewcommand{\thefootnote}{\arabic{footnote}}
\setcounter{footnote}{0}     %% Resetting the footnote counter 

%%%+++++++++++++++++++++++++++++++++++++++++++++++++++++++++++++++%%%
%%% Section : Introduction                                        %%%

\section{Introduction \label{section:1} }

One of the problems affecting the Standard Model (SM) of particle physics is the hierarchy
between the electroweak scale and the Planck scale.
Within the SM, the Higgs boson receives a quadratically divergent
contribution to its mass, if the model is considered an effective
theory, valid only up to some high energy scale. However, precision
electroweak measurements at colliders and at low energy clearly show that
the physics behind the SM is perturbative in nature. This means that
the Higgs boson mass cannot be very large and that the scale of new physics affecting those observables is larger than about 10 TeV, thus requiring fine-tuning between this new scale and the electroweak scale \cite{littlehierarchy}. In Little Higgs models (see
\cite{lhrev} for two recent reviews), the Higgs field is a Nambu-Goldstone boson
(NGB) of a global symmetry, which is spontaneously broken at some
higher scale $f$ by a vacuum expectation value. The Higgs
field gets a mass through symmetry breaking at the electroweak
scale. However, since it is protected by the approximate global
symmetry, it remains light. 
Generic Little Higgs models predict, at  
a scale of the order of $f$, new particles responsible for canceling
the SM quadratic divergences at the one-loop level: heavy $SU(2) \times U(1)$
gauge bosons, new heavy scalars and new fermions, in particular partners of the
top quark. The original Little Higgs models allowed
tree-level couplings of the new particles to the SM ones, inducing
tree  and loop level contributions to various electroweak precision
observables. Performing an analysis of the precision data for
generic regions of the parameter space, a lower bound on the symmetry breaking scale $f$
of several TeV was derived \cite{precisionEW},
reintroducing the fine tuning problem
between the cut-off scale of the models ($\sim 4 \pi f$) and the
electroweak scale. This resulted in addition in new Little Higgs particles that
were too heavy to be observed at the LHC.

In order to render this kind of models consistent with precision data and to
address the fine tuning problem, a new discrete parity named
$T$-parity (similar to $R$-parity in Supersymmetry) was introduced
\cite{tpar}. This new class of models, known as {\sl Little Higgs
  models with $T$-parity (LHT)}, includes new heavy particles, which are
postulated to be odd under $T$-parity. As a consequence, all the $T$-odd
particles can only be produced in pairs, so that they introduce no tree-level
contributions to electroweak observables. As the corrections to precision
observables now enter only at the loop level, they are naturally small.
The introduction of $T$-parity therefore allows to lower the new particle
mass scale to $f \sim 500$
GeV. In order to consistently implement $T$-parity, one has to
introduce sets of $T$-odd fermions corresponding to each of the SM
fermions. All the heavy particles introduced (except for a vector-like heavy top
quark $T_+$) are odd under $T$-parity.
These $T$-odd fermions have a mass of the order of $f$ and
can thus be abundantly produced at future colliders. 
LHT models also provide a possible candidate for dark matter 
that is odd under $T$-parity, namely a heavy photon $A_H$. 

Recently, the proton-proton Large Hadron Collider (LHC) has become
operational at CERN. One of the main focuses of this machine is the discovery of
the SM Higgs boson. In addition, it can probe 
various new physics models. In this work we will focus
on the possibility of the production of the first two generations of
$T$-odd quarks at the LHC. In particular we will focus on the dileptonic
signatures. 
The signatures we will study are (a) $ p p \to \ell^\pm
\ell^\mp j j \et$ and (b) $ p p \to \ell^\pm \ell^\pm j j \et$. The
former was also studied in \cite{Choudhury:2006mp}, where the authors
considered the pair production of $T$-odd quarks in the
channel $p p \to Q_H \bar{Q}_H$. They argued that this production
channel was QCD-dominated (with $q \bar{q}$ and $g g$ in the initial state),
and they neglected the electroweak (EW) contributions to it. In our study
we will show that the EW
contributions to these production channels are substantial. As
the EW contributions are not negligible, one can also have the
$T$-odd quark pair production via EW processes like $u u \to U_H U_H$
etc., giving a signature of same-sign dileptons in the final state. The
same-sign dilepton process has very small SM backgrounds and can therefore
be very useful for observing the LHT model at the LHC. In our analysis
we have used modified couplings of $T$-odd fermions, which
assure the correct cancellation of ultraviolet divergences in
$Z$-penguin diagrams in various flavour-changing decays \cite{Goto:2008fj}.

Our paper is organized as follows: In Sec.\ \ref{new}, we
present the novel aspects of our analysis. 
In Sec.\ \ref{section:2}, we
introduce briefly the particular LHT model that we have considered.
In Sec.\ \ref{section:3}, we discuss the production cross sections and
branching fractions of the pair production of the $T$-odd quarks and
their dependence on the LHT model parameters.  In Sec.\
\ref{section:4}, we describe our analysis set-up for both signal
and background processes. Finally, we conclude with a
summary of our results in Sec.\ \ref{section:5}. 

\vskip .3cm
%%%
\section{Overview \label{new}} 

The production and signatures of $T$-odd quarks at the LHC have previously been
discussed in Refs.\ \cite{Choudhury:2006mp,Freitas:2006vy,Belyaev:2006jh}.
In our analysis:
\begin{itemize}
\item{} we have included the $v^2/f^2$ corrections to the SM $Z$- and
  $W$-boson couplings to the mirror fermions as pointed out in Ref.\
  \cite{Goto:2008fj}\footnote{The revised CALCHEP model files that
  include the new corrections to the couplings of mirror quarks to
  $Z$- and $W$-bosons can be downloaded from 
  {\tt {http://deandrea.home.cern.ch/deandrea/LHTmodl.tgz}}.};

\item{} we have performed a detailed realistic simulation by using the fast
  detector simulator ATLFAST \cite{atlfast}
  for both signal and background processes;

\item{} we have considered the electroweak (EW) contributions to the
  production processes that have been neglected in Ref.\ \cite{Choudhury:2006mp};

\item{} we have analyzed same-sign dilepton signatures that had not been
  not considered in Refs.\ \cite{Choudhury:2006mp,Freitas:2006vy,Belyaev:2006jh};
  note that for these signatures the SM backgrounds are very small,
  so that this mode can be very useful to discover the $T$-odd quarks at the LHC;

\item{} we have given the $K$-factors for the production of $T$-odd quarks
  via QCD and EW diagrams, and we have included $K$-factors for both
  signal and backgrounds in our simulation results. 
\end{itemize}
\vfill

%%%+++++++++++++++++++++++++++++++++++++++++++++++++++++++++++++++%%%
%%% Section : Model                                               %%%

\section{The model \label{section:2}}

As an example for a typical spectrum of new particles introduced in
Little Higgs models with $T$-parity, we consider the Littlest Higgs
model with $T$-parity \cite{Low:2004xc,Hubisz:2004ft,Hubisz:2005tx}.  
We only briefly review here the aspects of the model relevant for our
analysis. 
The model we have considered has a $SU(5)$ global symmetry that is
broken down to $SO(5)$. The $[SU(2) \times U(1)]^2$ subgroup of $SU(5)$ is
gauged and is broken down to the diagonal subgroup $SU(2)_L \times U(1)_Y$,
that is identified with the SM electroweak gauge group. 
The masses of the heavy $T$-odd gauge bosons are
\begin{equation}
M_{A_H} \simeq \frac{g' f}{\sqrt{5}} , \ \ 
M_{V_H} \simeq g f , \ \
\end{equation}
where $g'$ and $g$ denote the hypercharge and $SU(2)$ weak couplings,
respectively. $A_H$ is usually the lightest $T$-odd particle\footnote{There
is a possibility of the $T$-odd neutrino to be a dark matter
candidate; this has been explored in \cite{Dey:2008dk}.}.

Concerning the implementation of $T$-parity in the fermion
sector, each SM fermion doublet is replaced by a pair of  
fields $F_i\, (i =1,2)$, where $F_i$ is a doublet under one $SU(2)_i$ and a
singlet under the other. $T$-parity exchanges $F_1 $ and $F_2$.
The $T$-even combination is  identified with the SM fermion
doublet, and the other ($T$-odd) combination 
is the heavy partner $F_H$. To generate mass terms for these $T$-odd
heavy fermions through Yukawa interactions, one requires additional $T$-odd 
$SU(2)$ singlet fermions in the theory. Assuming for simplicity a
universal and flavour diagonal Yukawa coupling $\kappa$, we have for
the heavy up-quark $U_H$ and the heavy down-quark $D_H$ (the $T$-odd heavy
partners of the SM quarks $(u,c)$ and $(d,s)$, respectively)
\begin{equation}
M_{D_{H,i}} \simeq \sqrt{2} \, \kappa_i \, f \ ,
\qquad
M_{U_{H_i}}  \simeq \sqrt{2} \, \kappa_i \, f \, 
         \left(1 - \, \frac{v^2_{\rm SM}}{8 \, f^2} \right)\ .
\end{equation}
The up- and down-type $T$-odd heavy quarks have nearly equal masses, as
the scale $f$ is typically at least in the 500 GeV range or more. The
Yukawa couplings $\kappa_i$ depend in general
on the fermion species $i$. This can in turn generate Flavour
Changing Neutral Current (FCNC) interactions in the quark sector
\cite{Blanke:2006eb}. Similar phenomena can also occur in the lepton
sector, giving rise to Lepton Flavour Violation (LFV) in this
class of models \cite{Goto:2008fj,Choudhury:2006sq}. For our analysis,
we will assume that the $\kappa_i$ are flavour-blind and universal and
hence do not give rise to any new sources of flavour violation. The
top sector requires typically an additional $T$-even fermion $T_+$ and a $T$-odd
$T_-$ to cancel the Higgs quadratic divergences. 
The cancellation of the quadratic divergence in the Higgs boson mass is not due to the $T$-odd
states, but it is achieved by loops involving the SM
top quark and the heavy $T$-even top quark. 
For a different implementation of the 
heavy top sector in $T$-parity models see Ref.\ \cite{Cheng:2005as}.

The LHT parameters relevant for our analysis are $f$ and $\kappa$. 
In our analysis we will assume that the values of
$\kappa$ are sufficiently smaller than the upper bound obtained from
four-fermion operators: $M_{\rm TeV} \leq 4.8
f^2_{\rm TeV}$ \cite{Hubisz:2005tx} where $M_{\rm TeV}$ and $f_{\rm TeV}$ 
are the $T$-odd fermion masses and symmetry breaking scale,
respectively, in TeV. For our analysis we have chosen three
representative points with values of $\kappa = 0.6, 1$ and $1.5$. 

%%%+++++++++++++++++++++++++++++++++++++++++++++++++++++++++++++++%%%
%%% Section : Production and decays of first two generations of   %%%
%%%           T-odd quarks at LHC                                 %%%

\section{\boldmath Production and decays of  the first two generations of $T$-odd
  quarks at the LHC  \label{section:3}}  

We compute the cross sections for the pair production of the first two generations
of $T$-odd quarks at the LHC and their branching fractions based on the model
briefly defined in Sec.\ \ref{section:2} and using CalcHEP 2.5.4
\cite{Pukhov:2004ca}. As described earlier, we have modified the
LHT model files for CalcHEP as proposed in Ref.\ \cite{Belyaev:2006jh} to
include the new $v^2/f^2$ contributions to the couplings of mirror
fermions to the SM $W$- and $Z$-bosons. All the cross sections are
calculated for a LHC centre-of-mass energy of 14 TeV. We have used the
leading order (LO) CTEQ6L parton densities with two-loop $\alpha_s$ and
$\Lambda_{\overline{\rm MS}}^{n_f=5}=226$ MeV \cite{Pumplin:2002vw} and
identified both the factorization scale $\mu_f$ and the renormalization scale
$\mu_r$ with the partonic centre-of-mass energy $\hat{s}$.

%%%>>>>>>>>>>>>>>>>>>>>>>>>>>>>>>>>>>>>>>>>>>>>>>>>>>>>>>>>>>>>>>>%%%
%%%>>>>>>>>>>>>>>>>>>>>>>>>>>>>>>>>>>>>>>>>>>>>>>>>>  Sub-section %%%  

\subsection{\boldmath Decays of first- and second-generation $T$-odd quarks}

\begin{figure}[htb]
\bcen
\hspace*{-2.5cm}
\epsfig{file=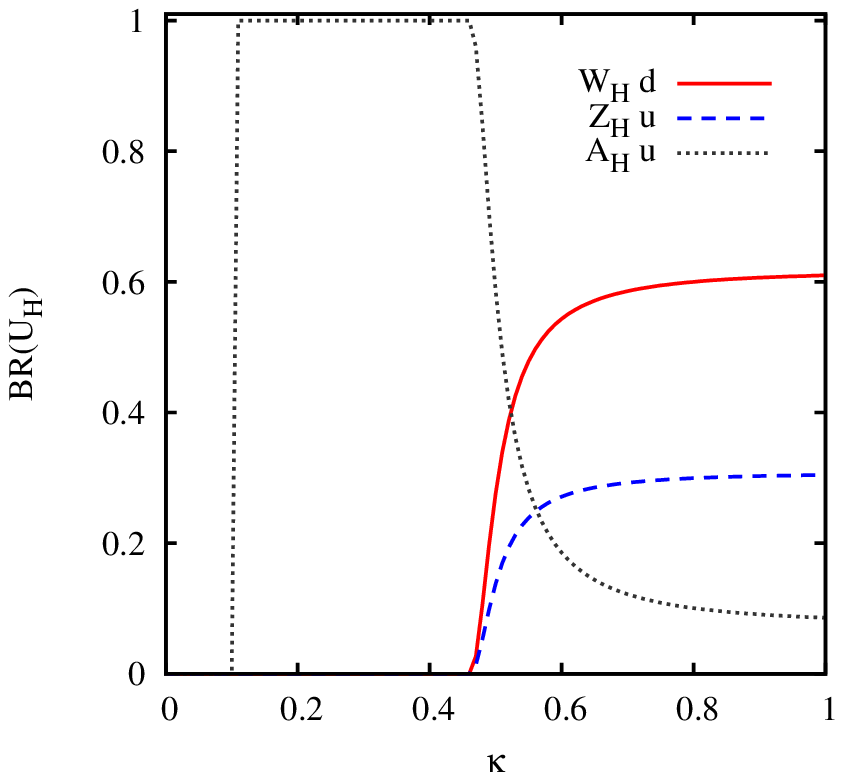,width=.62\textwidth} \hspace*{-3.2cm}
\epsfig{file=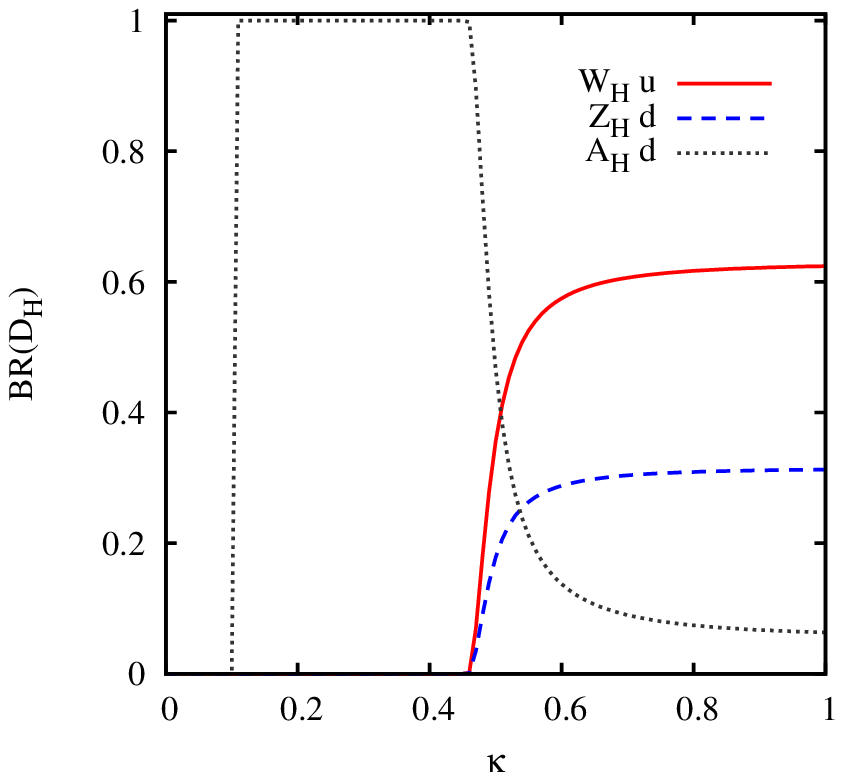,width=.62\textwidth} \hspace*{-2cm}
\vskip -.5cm
\caption{\sl Branching ratios of $T$-odd up-type ($U_H$, left panel) and down-type
 ($D_H$, right panel) heavy quarks as a function of $\kappa$ for a symmetry
 breaking scale of $f = 1000$ GeV.} 
\label{fig:br1}
\ecen
\vskip -.5cm
\end{figure}

The $T$-odd quarks will decay into a $T$-odd particle and a $T$-even
SM particle. The decay pattern is determined by the mass spectrum of
the $T$-odd particles. In LHT, we have typically
\begin{equation}
m_{A_H} \simeq \frac{g' f}{\sqrt{5}} \simeq 0.156 f\ , ~~~
m_{V_H} \simeq g f \simeq 0.653 f\ , ~~~
m_{Q_H} \simeq \sqrt{2} \kappa f \simeq 1.414 \kappa f\ ,
\end{equation}
where $m_{V_H}$ and $m_{Q_H}$ are the $T$-odd $W$-boson, $Z$-boson and quark
masses, respectively. We show the branching ratios of the up-type
($U_H$) and down type ($D_H$) $T$-odd quarks as a function of $\kappa$
in Fig.\ \ref{fig:br1}. 

%%=========================================%%
%% Figure 
%%=========================================%%

From the spectrum of $T$-odd particles, we can observe that $A_H$ is
always lighter than 
$W_H$. When $\kappa < 0.11$, one obtains a mass hierarchy $m_{Q_H} <
m_{A_H} < m_{W_H}$. In this case, the $T$-odd fermion would be a dark
matter candidate, but as discussed in Ref.\ \cite{Primack:1988zm}, the dark
matter  candidate should be a neutral and colourless object. So for our analysis
we will work in the parameter space where $\kappa > 0.11$. When
$ 0.11 < \kappa < 0.462$, the mass spectrum is $m_{A_H} < m_{Q_H}
< m_{W_H}$. In this case, the dominant decay mode of the $T$-odd heavy
quark is $Q_H \to A_H q$. If $\kappa > 0.462$, we have the mass
spectrum $m_{A_H} < m_{W_H} < m_{Q_H}$, and as shown in Fig.\
\ref{fig:br1}, the $T$-odd quark will predominantly decay through the
channel $Q_H \to W_H q'$.  In our analysis, we will restrict ourselves to the
parameter space $\kappa > 0.462$, where
BR$(W_H \to A_H W) = 100\, \%$.

If the masses of the
first two generations of $T$-odd quarks are not too high, they can
be produced in large numbers at the LHC. Assuming that the SM $W$-bosons decay
leptonically, the $T$-odd quarks for $\kappa > 0.46$ predominately
follow the decay chains
\beqa
U_H &\to& W_H^+ d \to W^+ A_H d \to \ell^+ \nu A_H d \to \ell^+ j \et,
\nonumber \\ 
D_H &\to& W_H^- u \to W^- A_H u \to \ell^- \bar{\nu} A_H u \to \ell^- j \et,
\nonumber \\ 
\bar{U}_H &\to& W_H^- \bar{d} \to W^- A_H \bar{d} \to \ell^- \bar{\nu} A_H
\bar{d} \to \ell^- j \et, \nonumber \\
\bar{D}_H &\to& W_H^+ \bar{u} \to W^+ A_H \bar{u} \to \ell^+ \nu A_H
\bar{u} \to \ell^+ j \et,
\label{decay-chain}
\eeqa
whose probability is approximately 12\% (taking into account the branching ratio of the $U_H$ and $D_H$ decays, and the leptonic decay of the $W$, where the lepton is either an electron or a muon).
In our analysis we will focus on the decay chains of $T$-odd heavy
quarks as listed above.

%%%                                               Sub-section END %%%  
%%%>>>>>>>>>>>>>>>>>>>>>>>>>>>>>>>>>>>>>>>>>>>>>>>>>>>>>>>>>>>>>>>%%%

%%%>>>>>>>>>>>>>>>>>>>>>>>>>>>>>>>>>>>>>>>>>>>>>>>>>>>>>>>>>>>>>>>%%%
%%%>>>>>>>>>>>>>>>>>>>>>>>>>>>>>>>>>>>>>>>>>>>>>>>>>  Sub-section %%%  

\subsection{Production cross sections at the LHC} 

\begin{figure}[htb]
\bcen
%\vskip -1cm
\vspace*{-2mm}
\hspace*{-8mm}
\epsfig{file=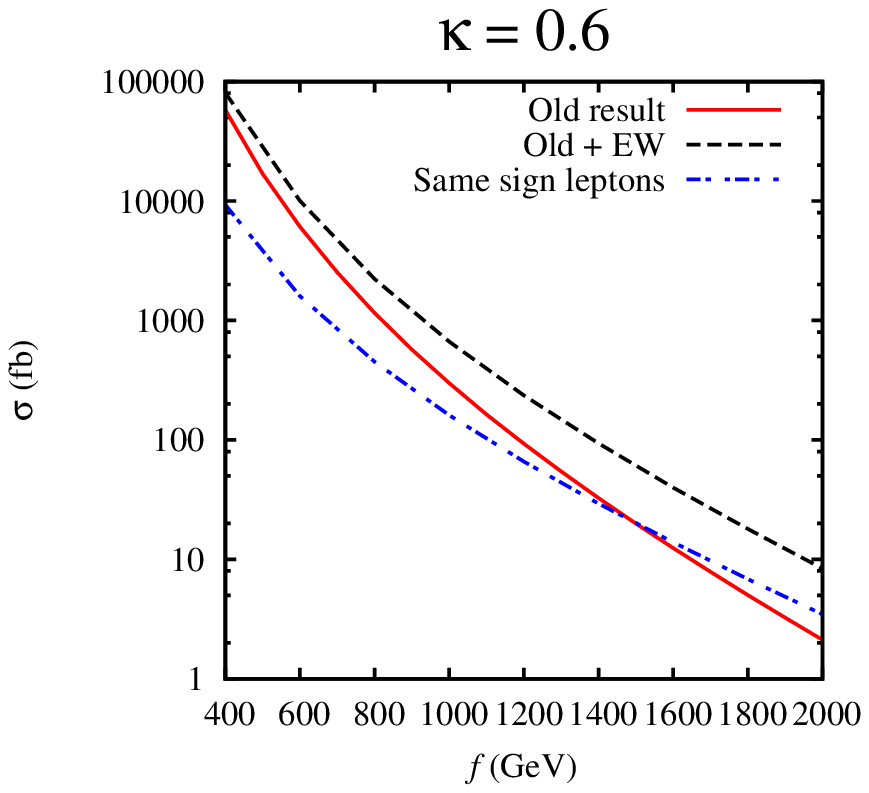,width=.6\textwidth} \hspace*{-3.2cm}
\epsfig{file=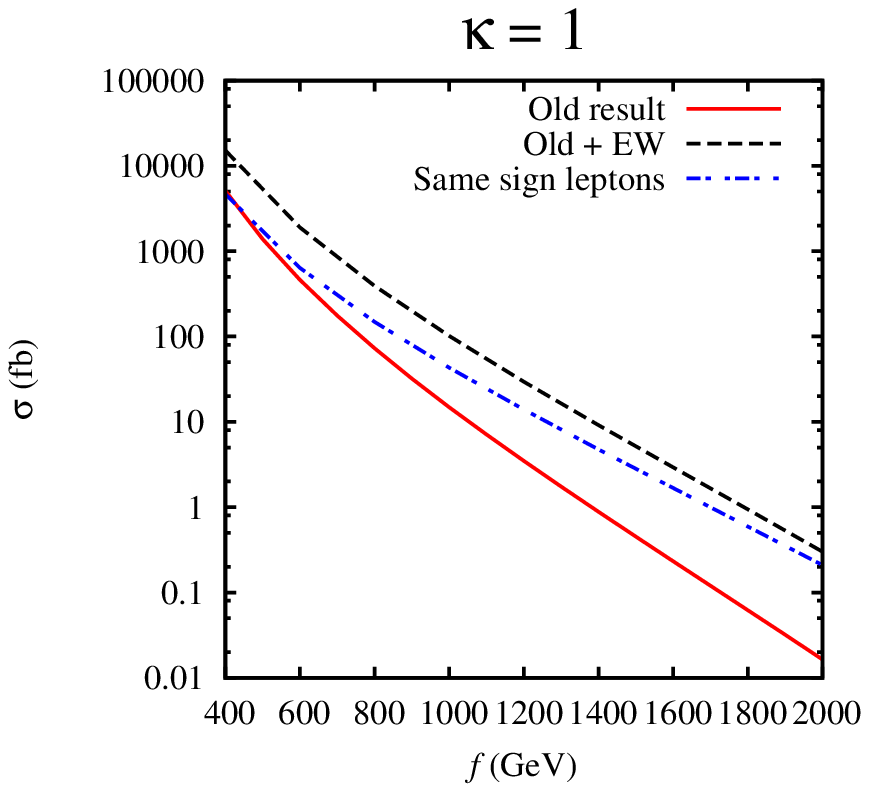,width=.6\textwidth} \hspace*{-3.2cm}\\
\vspace*{-8mm}
\epsfig{file=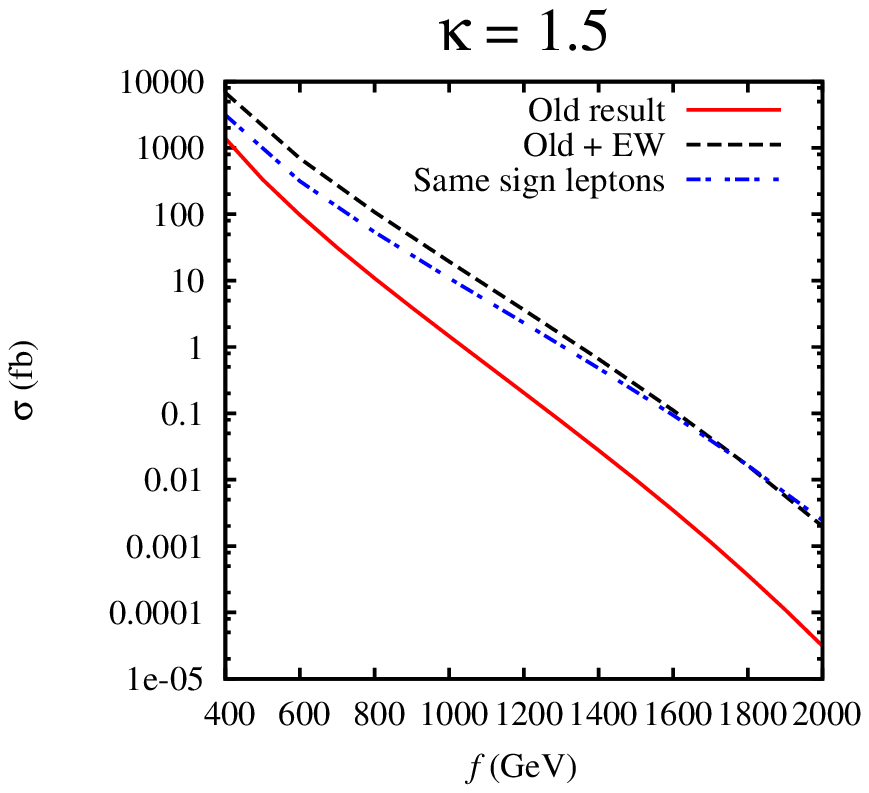,width=.6\textwidth}
\vspace*{-8mm}
\caption{\sl Production cross sections for a pair of
 $T$-odd quarks at the LHC with a centre-of-mass energy of 14 TeV.
The plots include all the channels in Eqs.\  (\ref{eq:prod:1}) and
(\ref{eq:prod:2}). The label ``Old result'' corresponds to pure QCD
contributions \cite{Choudhury:2006mp}, while the label ``Old + EW''
represents the complete set of QCD + EW diagrams. The label ``Same sign
leptons'' marks the cross sections for the EW channels in
Eq.\ (\ref{eq:prod:3}), which contribute to the same-sign lepton
signature.} 
\label{fig:cs1}
\ecen
\end{figure}

The pair production of $T$-odd quarks at the LHC has been considered in
Refs.\ \cite{Choudhury:2006mp,Freitas:2006vy,Belyaev:2006jh}. 
A detailed signal and background estimation of the pair production
of $T$-odd quarks was carried out by Choudhury {\sl et al.}
\cite{Choudhury:2006mp} using a parton-level Monte Carlo generator, where the
authors considered in particular the $\ell^\pm \ell^\mp j j \et$ signature. 
This signature is generated
by the following production channels:
\beqa
p p &\to& Q_H \bar{Q}_H,~~~~~~~~~~  {\rm with} ~~ Q_H = U_H, D_H, C_H, S_H  \
\ \,  ({\rm QCD} + {\rm EW})\, ,  \label{eq:prod:1} \\
p p &\to& Q_H Q'_H + c.c. , ~~ {\rm with} ~~ Q_H = U_H, C_H, ~~ Q'_H =
D_H, S_H \ \ \ ({\rm EW})\,. \label{eq:prod:2}
\eeqa
Choudhury {\sl et al.} considered only the QCD part of the
production channels given in Eq.\ (\ref{eq:prod:1}), arguing that the
EW amplitudes would be much smaller than those mediated by QCD. 
We have 
evaluated the production cross sections of both
channels given in Eqs.\ (\ref{eq:prod:1}) and (\ref{eq:prod:2}).
Our results are shown in Fig.\ \ref{fig:cs1}. For comparison with the
results in literature, we have chosen the same set of
input parameters as in Fig.\ 1 of Choudhury {\sl et al.}
\cite{Choudhury:2006mp}.  As it can be seen from our figure, the EW
contribution can substantially alter the QCD results, and in some cases
the full cross section is
enhanced by one order of magnitude.

%%=========================================%%
%% Figure 
%%=========================================%%

The reason why the EW contribution is in this case substantial is the following:
the EW contributions are basically $t$-channel diagrams with $1/(t-m^2)$
dependence, which leads to large contributions in the near-forward region of
$\cos\theta = 1$. Here, $t$ is the invariant momentum transfer squared, $m$ is
the mass of the exchanged particle, and $\theta$ is the angle between the initial
and final state particles in the centre-of-mass frame. This behaviour is very
different from the $s$-channel diagrams, which dominate the QCD contributions
and have a regular $(1+\cos^2\theta)$ dependence. If the particle masses and
couplings in a particular model are suitably chosen, then it is indeed possible
for $s$-channel diagrams to fall off more rapidly with $1/s$ than the $t$-channel
diagrams. This is exactly what is happening in the LHT model considered here,
despite the suppression due to the weak couplings.

Figure \ref{fig:cs1} clearly indicates that EW diagrams,
via $t$-channel exchange, can be comparable to $s$-channel
QCD contributions. It is then interesting to consider purely weak
processes like
\beq
 p p \to Q_H \bar{Q'}_H + c.c. , ~~~
{\rm with} ~ Q_H = U_H, C_H , ~~ Q'_H = D_H, S_H. 
\label{eq:prod:3} 
\eeq
These channels, following the
decay chains in Eqs.\ (\ref{decay-chain}), give rise to same-sign
dilepton signatures, $\ell^\pm \ell^\pm j j \et$, with two same-sign
leptons, two light jets, and missing transverse energy $E_T$. Same-sign
dileptons have relatively small SM backgrounds and are therefore very
distinctive for physics beyond the SM. In Fig.\
\ref{fig:cs1}, we have also shown the production cross section for the
processes given in Eq.\ (\ref{eq:prod:3}).

%%=========================================%%
%% Figure 
\begin{figure}[htb]
\bcen
%\vskip -1cm
\vspace*{-2mm}
\hspace*{-8mm}
\epsfig{file=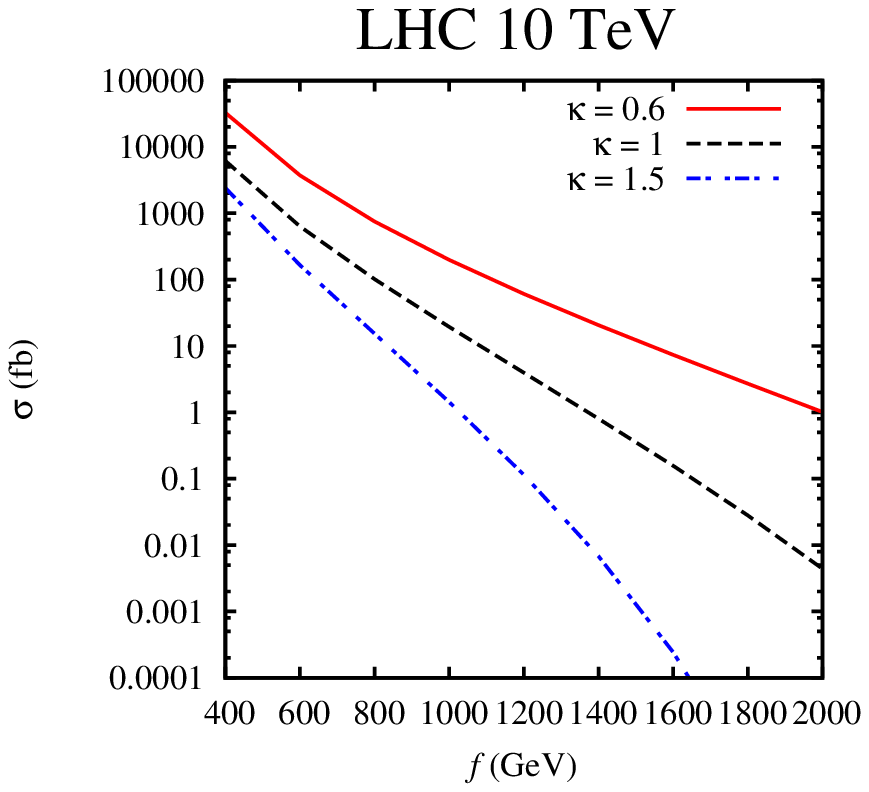,width=.6\textwidth} \hspace*{-3.2cm}
\epsfig{file=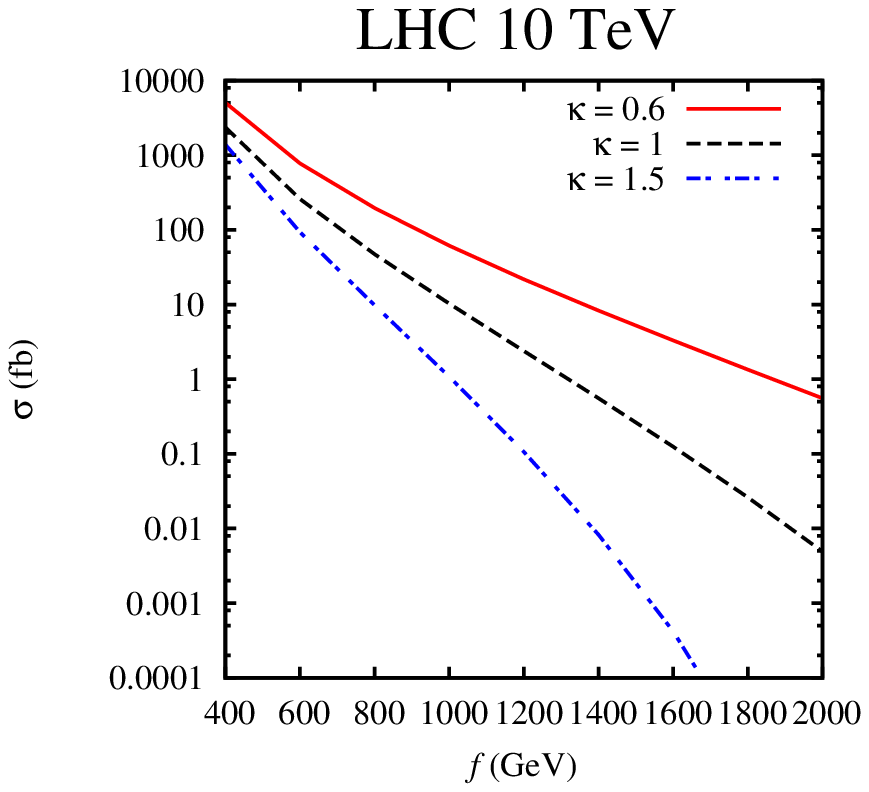,width=.6\textwidth} \hspace*{-3.2cm}\\
\vspace*{-8mm}
\caption{\sl Production cross sections at the LHC with a centre-of-mass energy of
 10 TeV for different-sign dilepton channels (left panel) and same-sign dilepton
 channels (right panel).}
\label{fig:cs2}
\ecen
\end{figure}
%%=========================================%%

As can be seen from Fig.\ \ref{fig:cs1}, the production cross sections
for processes giving rise to opposite-sign dileptons ($\ell^\pm \ell^\mp j j \et$) could be of
the the order of a picobarn, and those contributing to same-sign
dileptons ($\ell^\pm \ell^\pm j j \et$) could be of the order of few hundred
femtobarn for a reasonable
range of allowed parameter space in our LHT model.
At the moment, the program for the early LHC running consists of a
first run for $\sqrt{s} = 7$ TeV at very low luminosity ($< 100$
pb$^{-1}$), followed by an upgrade to $\sqrt{s} \sim 10$ TeV with a
luminosity of ${\cal L} \sim {\cal O}(100)$ pb$^{-1}$ \cite{early-lhc}.  
In Fig.\ \ref{fig:cs2}, we therefore also show the cross sections of the two
channels at the LHC with a centre-of-mass energy of 10 TeV. It can be seen
that even in very early stages and with a lower energy of 10 TeV it is
possible to copiously produce large number of $T$-odd quarks in a
reasonable range of LHT model parameter space. 
However, for our simulations we will assume an LHC run
at 14 TeV. Armed with the production cross sections and decay rates
for the signal processes, we will discuss in next section in detail the
signal and background event rates.  

\pagebreak

%%%>>>>>>>>>>>>>>>>>>>>>>>>>>>>>>>>>>>>>>>>>>>>>>>>>>>>>>>>>>>>>>>%%%
%%%>>>>>>>>>>>>>>>>>>>>>>>>>>>>>>>>>>>>>>>>>>>>>>>>>  Sub-section %%%  

%%%+++++++++++++++++++++++++++++++++++++++++++++++++++++++++++++++%%%
%%% Section : Signal and Background estimates                     %%%

\section{Signal and background estimates \label{section:4}}

In this section we will compute estimates for the signal and backgrounds
events. We will also study various kinematical distributions
for both signal and backgrounds that can be useful in extracting the
signal from the backgrounds. 
In our analysis, we have chosen three LHT model parameter space
points. The mass spectra of LHT particles relevant for our study are
given in Tab.\ \ref{table:spectra}.

%%+++++++++++++++++++++++++++++++++++++++++++++++++++++++++++++++
%% Table 
\begin{table}[htb]
\bcen
\caption{\sl Input parameters and masses of the LHT particles used for our
  analysis. } 
\label{table:spectra}
\vspace*{5mm}
\begin{tabular}{| c || c | c | c |} \hline 
Model parameters $\rightarrow$ & 
$f = 1000$ GeV & $f = 1000$ GeV & $f = 700$ GeV  \\ 
Particle masses (in GeV) $\downarrow$ &  
$\kappa = 0.6$ & $\kappa = 1$ & $\kappa = 1.5$ \\ 
\hline 
$M_{A_H}$ & 150 & 150 & 100 \\
$ M_{V_H}$ & 648 & 648  & 450 \\
$M_{U_H}$ &  842 & 1403  & 1462 \\
$M_{D_H}$ &  848 & 1414  & 1484 \\
 \hline 
\end{tabular}
\ecen
\end{table}
%%+++++++++++++++++++++++++++++++++++++++++++++++++++++++++++++++
\vspace*{-10mm}

%%%>>>>>>>>>>>>>>>>>>>>>>>>>>>>>>>>>>>>>>>>>>>>>>>>>>>>>>>>>>>>>>>%%%
%%%>>>>>>>>>>>>>>>>>>>>>>>>>>>>>>>>>>>>>>>>>>>>>>>>>  Sub-section %%%  

\subsection{Framework for event generation} 
The set-up used for signal and background generation is the following:
\begin{itemize}
\item{} {\bf Signal event generation:} We have used CalcHEP 2.5.4
  \cite{Pukhov:2004ca} to
  calculate cross sections and branching ratios. As described earlier,
  we have modified the LHT model file given by the authors in Ref.\
  \cite{Belyaev:2006jh}. The original LHT model file for CalcHEP \cite{Belyaev:2006jh}
  does not have a QNUMBERS block for the new particles, although the new
  version of CalcHEP (2.5.4) is compatible with this block. Apart from
  other modifications described earlier, we have 
  also introduced this block and accordingly passed the Monte Carlo
  numbers to the new set of particles in the revised model files. The
  parton level events generated by CalcHEP 2.5.4 were then passed
  on to PYTHIA 6.4.21 \cite{Sjostrand:2006za} via the LHE (Les
  Houches Event) interface \cite{Alwall:2007mw} in order to include
  initial and final state radiation (ISR/FSR) effects.
\item{} {\bf Background event generation:} We have generated the $t
  \bar{t}$ backgrounds using PYTHIA 6.4.21, still including ISR/FSR
  effects. The $W^\pm W^\mp j j$, $W^\pm W^\pm j j$, $Z Z
  jj$ backgrounds were generated using MADGRAPH
  \cite{Maltoni:2002qb} and were then 
  passed on to PYTHIA 6.4.21 for ISR/FSR effects. 
\end{itemize}

The $K$-factors for both signal and background processes have been computed with
MCFM 5.6 \cite{mcfm}, using for the LO cross sections the same LO parton
densities as in the full simulation, i.e.\ CTEQ6L with two-loop $\alpha_s$ and
$\Lambda_{\overline{\rm MS}}^{n_f=5}=226$ MeV \cite{Pumplin:2002vw}. For the
next-to-leading order (NLO) cross sections, we used the most recent NLO parton
densities CTEQ6.6M with an improved treatment of heavy-quark effects through a
general-mass variable-flavor number scheme \cite{Nadolsky:2008zw}. In both cases,
we identified the factorization scale $\mu_f$ and the renormalization scale
$\mu_r$ with the partonic centre-of-mass energy $\hat{s}$.

To be specific, the $K$-factors for the QCD signal cross section were computed
from $t\bar{t}$ production by increasing the top-quark mass from 175 GeV to
generic heavy-quark mass values of up to 2 TeV (see the second line in Tab.\
\ref{tab:1}). This
\begin{table}
\begin{center}
\caption{\label{tab:1}\sl Applied $K$-factors for the QCD and EW
 contributions to the signal process as a function of the heavy-quark mass
 $m_{Q_H}$.}
\vspace*{5mm}
\begin{tabular}{|c|cccccccccc|}
\hline
 $m_{Q_H}/$GeV  & 200  &  400 &  600 &  800 & 1000 & 1200 & 1400 & 1600 & 1800 & 2000 \\ % & 2200 & 2400 \\
\hline
 QCD $K$-factor & 1.20 & 1.43 & 1.51 & 1.54 & 1.56 & 1.58 & 1.58 & 1.52 & 1.46 & 1.42 \\ % & 1.40 & 1.37 \\
 EW  $K$-factor & 1.03 & 1.17 & 1.29 & 1.41 & 1.53 & 1.68 & 1.84 & 2.01 & 2.24 & 2.39 \\ % & 1.   & 1.   \\
\hline
\end{tabular}
\end{center}
\end{table}
is possible, since the QCD properties of $T$-odd quarks are identical to those
of the SM heavy quarks. Using the same method and similar parton densities, but
setting the scales to the heavy-quark mass, results at NLO+NLL (next-to-leading
logarithmic level) have been obtained for the LHC in Tab.\ 4 of Ref.\
\cite{Cacciari:2008zb}. Since NLL results are not available for EW production and
the background processes and in order to employ the same scales for all signal and
background processes, we do not make use of these cross sections here. The NLO+NLL
cross sections are moreover considerably higher than ours, so that our estimate of
corrected signal cross sections and consequently of the significance is quite
conservative.

$K$-factors for the electroweak signal cross sections cannot be computed with
MCFM, since the $s$-channel single-top production process $q\bar{q}'\to t\bar{b}$
was computed there with $m_b=0$. Instead, we take them from the $t$-channel
process $qg\to qt'\bar{b}'$ with $m_{b'}=m_{t'}$ as tabulated in Tab.\ 10 of Ref.\
\cite{Campbell:2009gj}, averaging over opposite charges of the heavy-quark
final state (see the third line in Tab.\ \ref{tab:1}).
The QCD and EW $K$-factors as a function of the heavy-quark mass in Tab.\
\ref{tab:1} lend themselves to quadratic and linear fits, respectively,
\bea
 K_{\rm QCD}~=~ 1.58-\lr{m_{Q_H}/{\rm GeV}-1200\over2000}\rr^2&,& K_{\rm EW}~=~{m_{Q_H}/{\rm GeV} + 1200\over1400},
\eea
which we have used in our full signal simulation.

The $K$-factors for the QCD and EW production of the SM $t\bar{t}$ background
have been computed for $m_t=175$ GeV as described above and are tabulated in the
second and third columns of Tab.\ \ref{tab:2}. However, the $K$-factors for the
\begin{table}
\begin{center}
\caption{\label{tab:2}\sl Applied $K$-factors for the considered background
 processes.}
\vspace*{5mm}
\begin{tabular}{|c|ccccc|}
\hline
 Final state & $t\bar{t}$ (QCD) & $t\bar{t}$ (EW) & $W^+jj$ & $W^-jj$ & $Zjj$ \\
\hline
 $K$-factor  & 1.16             & 1.01            & 1.52    & 1.47    & 1.32  \\
\hline
\end{tabular}
\end{center}
\end{table}
$WW+2$ jet and $ZZ+2$ jet backgrounds are not yet available in the literature.
We therefore computed with MCFM the $W+2$ jet and $Z+2$ jet $K$-factors
and assumed that the second vector boson, being uncharged under color, does not
change the $K$-factors significantly. For these backgrounds, we implemented all
the kinematic cuts described below. In particular, the jets were identified with
the midpoint cone algorithm of radius $R=0.7$ and parton separation $R_{\rm sep}
=1.3R$. The $Z+2$ jet background required in addition a minimum invariant mass cut
for the charged-lepton pair, which we have set to 15 GeV as in Ref.\
\cite{Campbell:2002tg}. Our results are listed in columns four to six of Tab.\
\ref{tab:2}.

In order to make realistic estimates of the signal and backgrounds, we
have further processed both the signal and background events through
the fast ATLAS detector simulator ALTFAST \cite{atlfast}. The resulting events have been analysed
within the ROOT framework. The detector simulator ATLFAST
provides a simple detector simulation and jet reconstruction using a
simple cone algorithm. It also identifies isolated leptons, photons,
b and $\tau$ jets and also reconstructs the missing energy. In our
analysis, leptons means electrons or muons {\sl i.e.} $\ell = e, 
\mu$. As stated above, we have assumed an LHC with a centre-of-mass energy
of $14$ GeV and luminosity ${\cal L} = 100$ fb$^{-1}$.

%%%>>>>>>>>>>>>>>>>>>>>>>>>>>>>>>>>>>>>>>>>>>>>>>>>>>>>>>>>>>>>>>>%%%
%%%>>>>>>>>>>>>>>>>>>>>>>>>>>>>>>>>>>>>>>>>>>>>>>>>>  Sub-section %%%  
\subsection{Opposite-sign dilepton signatures: $\signatureone$}

This signature is generated by the production processes given in Eqs.\
(\ref{eq:prod:1}) and (\ref{eq:prod:2}) and the decay chains given in Eqs.\
(\ref{decay-chain}).

%%=========================================%%
%% Figure 
\begin{figure}[tb]
\bcen
\epsfig{file=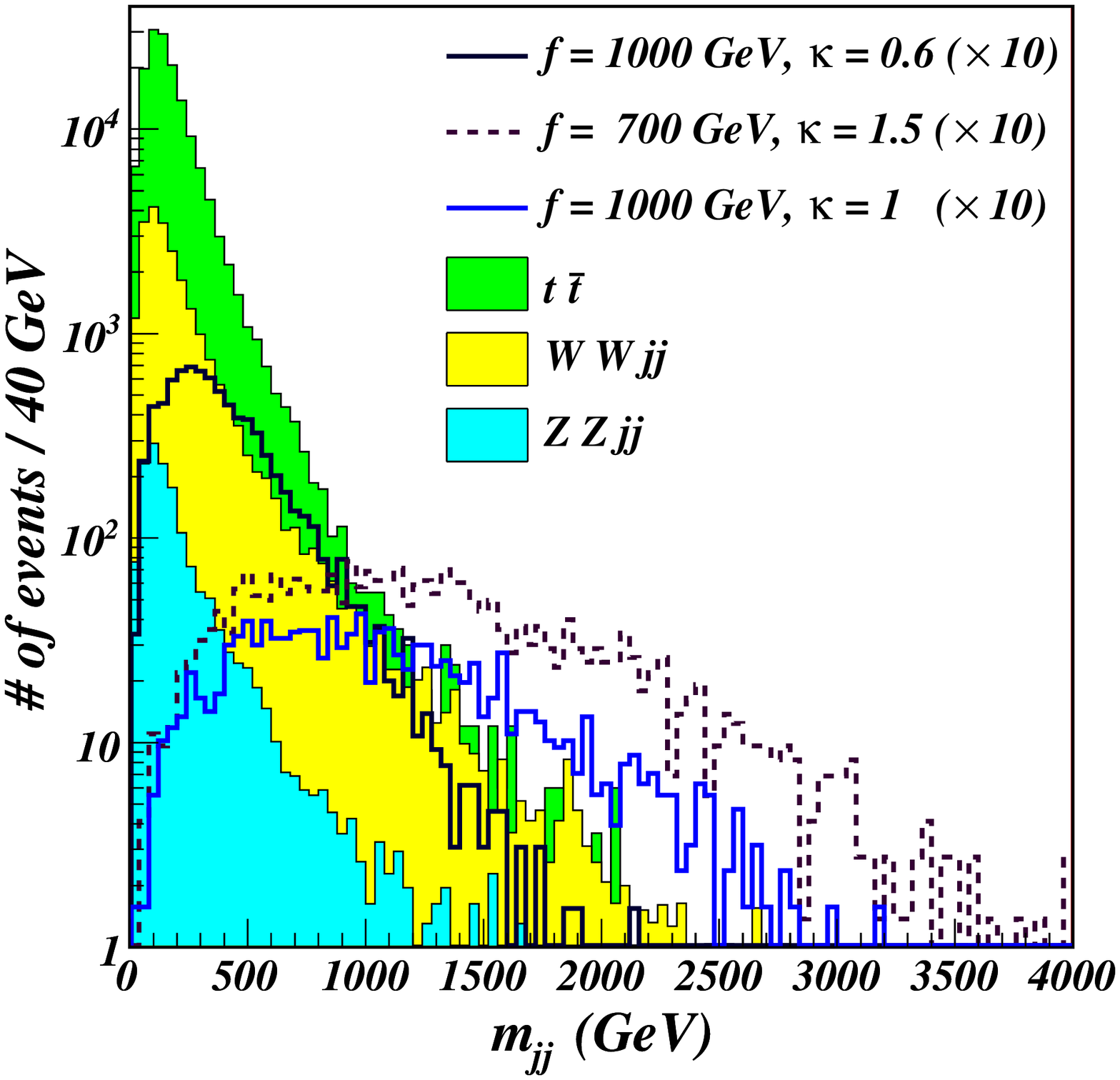,width=.5\textwidth} \hspace{-.5cm}
\epsfig{file=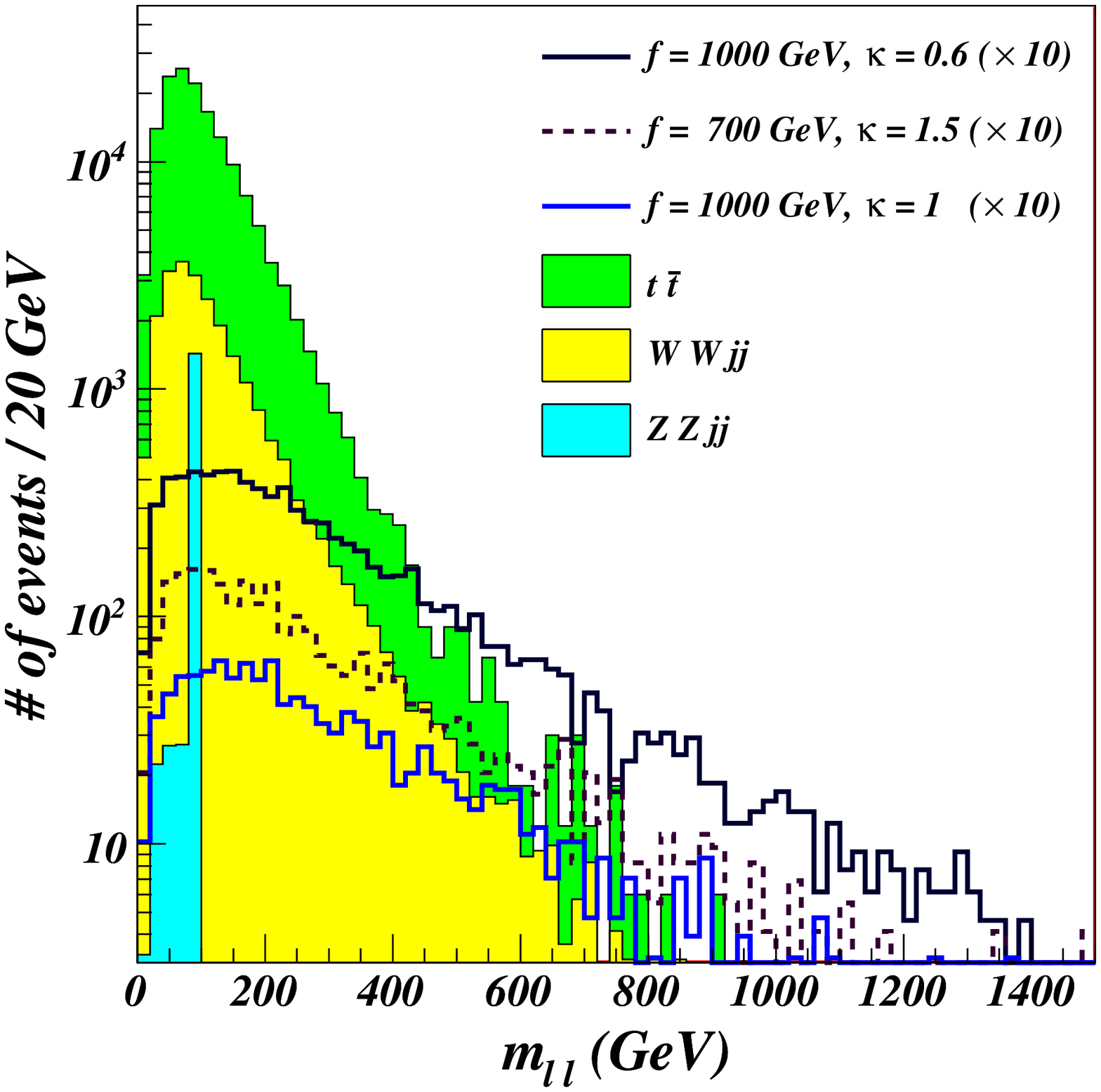,width=.5\textwidth} 
\vspace*{-.2cm}
\caption{\sl Dijet invariant mass ($m_{jj}$) distribution (left panel)
  and di-lepton invariant mass ($m_{\ell^\pm \ell^\mp}$) distribution  
  (right panel) for signal ($\times 10$) and SM background in the
  channel $\pptoqqbjjll$.  In plotting these
  distributions we have assumed the LHC luminosity to be ${\cal L} = 100$
  fb$^{-1}$. }
\vspace{-.3cm}
\label{fig:todd:1}
\ecen
\end{figure}
%%=========================================%%
%%=========================================%%
\begin{figure}[tb]
\bcen
\epsfig{file=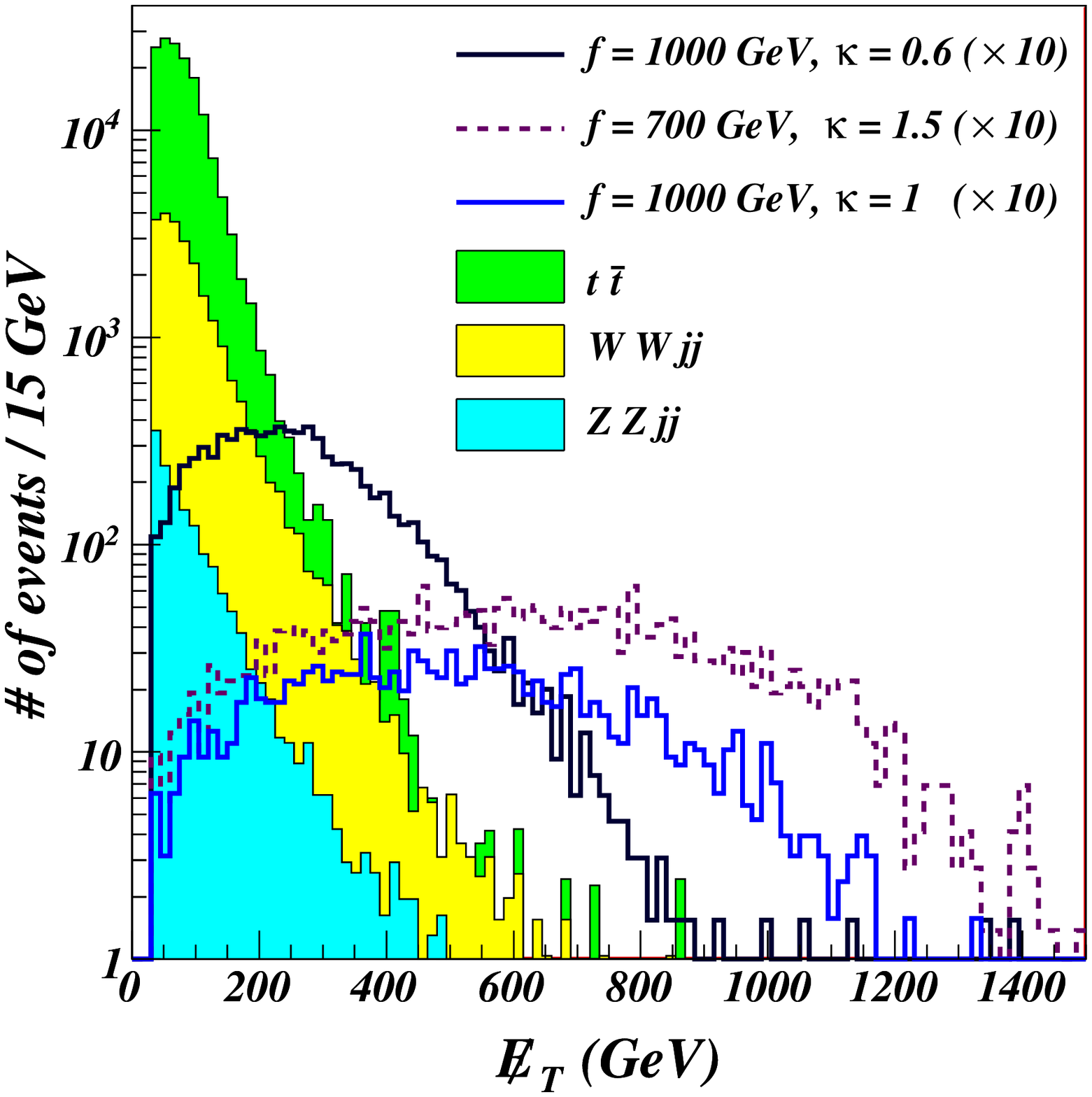,width=.51\textwidth} \hspace{-.5cm} 
\epsfig{file=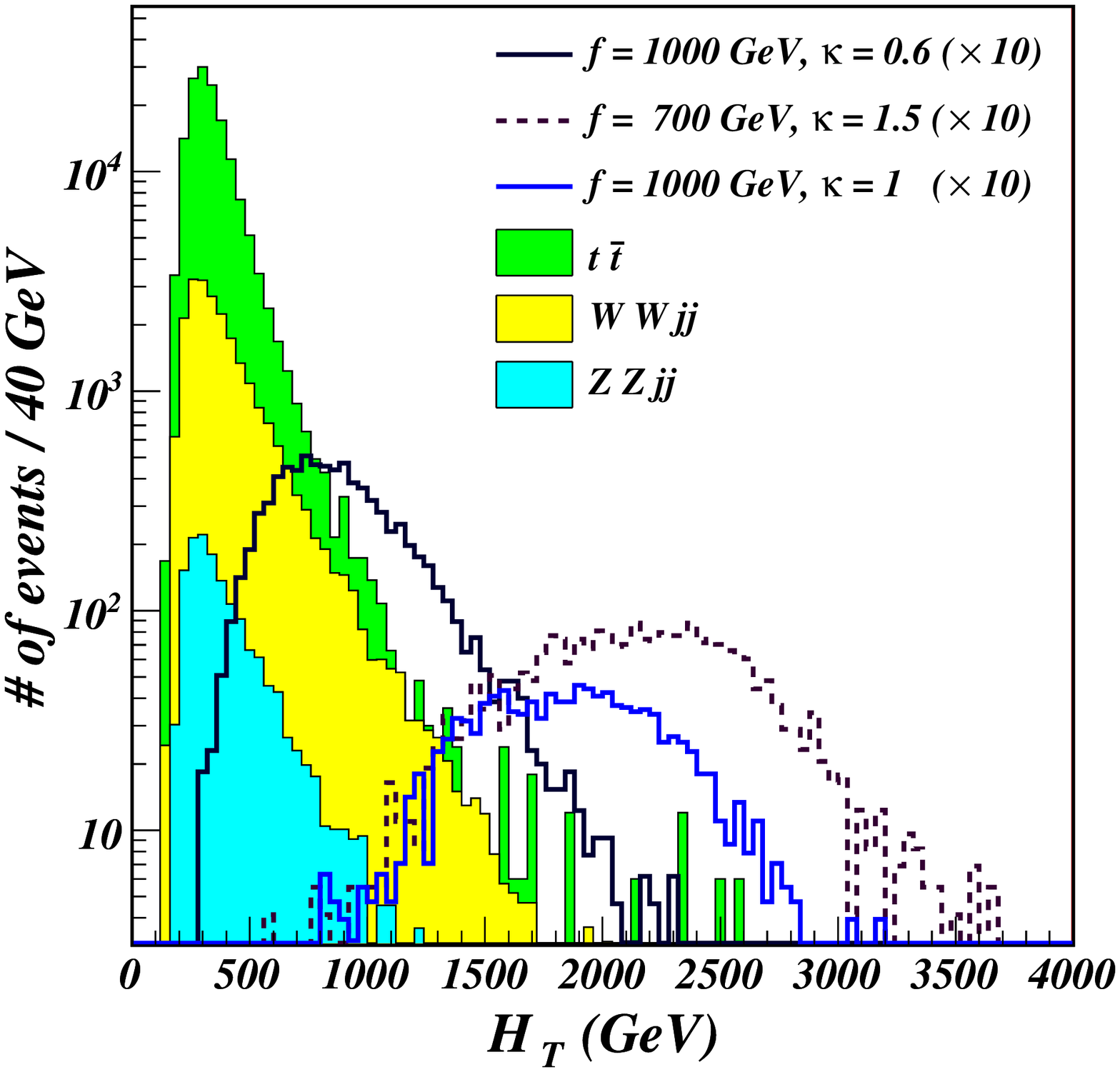,width=.5\textwidth}
\vspace*{-.2cm}
\caption{\sl $\et$ and $H_T$ distributions for  signal ($\times 10$) and SM
  background in the channel $\pptoqqbjjll$.  In plotting these 
  distributions we have assumed the LHC luminosity to be ${\cal L} = 100$
  fb$^{-1}$. }
\vspace{-.3cm}
\label{fig:todd:2}
\ecen
\end{figure}
%%=========================================%%

The possible SM background sources to this signal are:
\begin{itemize}
\item{} $p p \to t \bar{t}$  with each of the two top quarks
  decaying via $t \to b W (\to \ell \nu) \to b \ell \et$ and the
  $b$-jet misidentified as a light jet. 
\item{} $p p \to W^+ W^- j j$ with both $W$-bosons decaying
  leptonically through $W \to \ell \nu$. 
\item{} $p p \to Z Z j j$ with one of the $Z$-bosons decaying
  via $Z \to \ell^+ \ell^-$ and the other $Z$-boson decaying to neutrinos
  $Z \to \nu \bar{\nu}$. 
\end{itemize}

\noindent In order to study the signal and background, we have implemented the
following {\sl pre-selection} cuts :
\begin{itemize}
\item[(a)] exactly two opposite charge leptons ($e, \mu$) in the event with
  $p_T^\ell > 15$ GeV and rapidity in the range $|\eta| < 2.5$;
\item[(b)] $b$-jet veto (reject any event having a well identified
  $b$-jet); we have considered a $b$-tagging efficiency of 60\%; this
  cut helps in reducing the $t \bar{t}$ background;
\item[(c)] exactly two light jets with $p_T^j > 30$ GeV and $|\eta| <
  2.5$;
\item[(d)] minimum missing energy threshold: $\et > 30$ GeV;
\item[(e)] minimum threshold for the opposite-sign dilepton invariant mass
  $m_{\ell^+ \ell^-} > 15$ GeV; this helps in reducing the backgrounds
  where the lepton pair originates from a virtual photon. 
\end{itemize}

We have shown the dijet (for lighter jets) invariant mass distribution
($m_{jj}$) and the opposite-sign dilepton invariant mass distribution
($m_{\ell \ell}$) for signal and backgrounds in Fig.\ \ref{fig:todd:1}. 
As it can be observed from these distributions, 
we can reduce the backgrounds, where the lepton pair and the jet pair
originate from a $Z$- or $W$-boson. For this we impose the extra condition
that the invariant mass of jets and leptons are away from $M_Z$ and
$M_W$, respectively. So we demand:
\begin{equation}
m_{jj} ~ \notin ~ [65,105] ~ GeV \ , ~~~ m_{\ell \ell} ~ \notin ~
[75,105] ~ GeV \ .
\end{equation}

Tab.\ \ref{table:tq:1} summarizes our results in the channel
$\signatureone$. In this table, we have shown the incremental effects on
the cuts defined above on signal and background. As it can be seen from
the table, the signal events are not much affected by the imposition of
the cuts on the dijet and dilepton invariant masses, whereas we can
substantially reduce the backgrounds arising from the $W W jj$ and $Z Z
jj$ channels.

We can also use the $\et$ cut to further reduce the backgrounds. The $\et$ cut is very 
useful, because in the SM, $\et$ originates from neutrinos that come from
the decay of $W$ or $Z$, and hence it could be relatively soft, whereas in the
LHT model, $ \et$ comes from a heavy particle, the $T$-odd photon ($A_H$),
and hence it could be relatively hard. We have shown the $\et$
distribution in Fig.\ \ref{fig:todd:2}, where one can see that the background
can be reduced by using a harder $\et$
cut. We have accordingly shown the results in Tab.\ \ref{table:tq:1}
by using cuts $\et > 200, 300, 400$ GeV

%%+++++++++++++++++++++++++++++++++++++++++++++++++++++++++++++++
%% Table 
\begin{table}[htb]
\bcen
\caption{\sl The $\sigma$ numbers shown in the second row are with
  relevant $K$-factors included. The $\sigma$ numbers in brackets
  (second row) are the cross section values, if we only include the QCD
  production mechanism without $K$-factors as considered in Choudhury
  et al. \cite{Choudhury:2006mp} as given in
  Eq.\ \ref{eq:prod:1}. The remaining rows indicate the number 
  of events for the luminosity ${\cal L} = 100$ fb$^{-1}$. For the
  numbers given above, we have included both QCD and EW
  contributions. } 
\label{table:tq:1}
\begin{tabular}{| c | c | c | c || c | c | c |} \hline 
Parameter set $\Rightarrow$ & $f = 1000$ &  $f = 1000$  & $f
= 700$ & SM & SM & SM \\
Cuts $\Downarrow$  & $\kappa = 0.6$ & $\kappa = 1$ & $\kappa = 1.5$ & $t \bar{t}$ & $W^+ W^-jj$  & $ZZjj$    \\ \hline 
Production $\sigma$ (fb)    & 1039.1 (298) &  157.4 (14.8)  & 412.3 (31.4)  &    &     &           \\
Preselection cuts             & 795.7     & 120.7  & 262.6 & $1.54 \times 10^5$ & $2.29 \times 10^4$  & 1520.6  \\
$m_{jj} \notin [65,105]$    & 755    & 120.1  & 261.7 & $1.26
\times 10^5$ & $1.88 \times 10^4$  &  1227.5  \\
$m_{\ell \ell} \notin [75,105]$  & 696.8 & 111.9  & 239.4 & $9.94 \times 10^4$ & $1.5 \times 10^4$  &  64.5   \\
$\et >$ 100                 &   623.8   & 108.8  & 234.5 & $2.5 \times 10^4$ & 4946.4 &  19.9     \\
$\et >$ 200                 &   441.2   & 100.5  & 220.3 & 2136.4 & 899.5 &  3.9     \\
$\et >$ 300                 &   237.3   & 87.5   & 200.1 &  396.1 & 239   &  1.3     \\
$\et >$ 400                 &   107.1   & 71.9   & 174.6 &  114.1 & 69.5  &  0.7     \\   \hline 
${\cal S}$                  &   7.2     &  4.9   &  11.2 &   &    &   \\ \hline 
\end{tabular}
\ecen
\end{table}
%%+++++++++++++++++++++++++++++++++++++++++++++++++++++++++++++++

In addition, one can also use $H_T$ (the total transverse energy) 
as the parameter to distinguish signal and backgrounds.  
The energy of the heavy particles produced ($T$-odd quarks for signal) is
essentially given to its various products, namely jets ($j$), leptons
($\ell$) and $\et$. Therefore one can define the total transverse energy
($H_T$) as :
\beq
H_T = \sum_{j,~ \ell, ~~ \et} |\vec{p}_T|. 
\eeq
The $H_T$ distribution peaks around the heavy particle mass, and a cut on
$H_T$ could be helpful in reducing the backgrounds. 
The $H_T$ distribution is shown in Fig.\ \ref{fig:todd:2}. As the $H_T$
distribution tends to peak around the heavy particle mass, it can
also be used to estimate the $T$-odd quark masses.

%%%>>>>>>>>>>>>>>>>>>>>>>>>>>>>>>>>>>>>>>>>>>>>>>>>>>>>>>>>>>>>>>>%%%
%%%>>>>>>>>>>>>>>>>>>>>>>>>>>>>>>>>>>>>>>>>>>>>>>>>>  Sub-section %%%  
\subsection{Same-sign dilepton signatures: $\signaturetwo$} 

%%=========================================%%
%% Figure 
\begin{figure}[tb]
\bcen
\epsfig{file=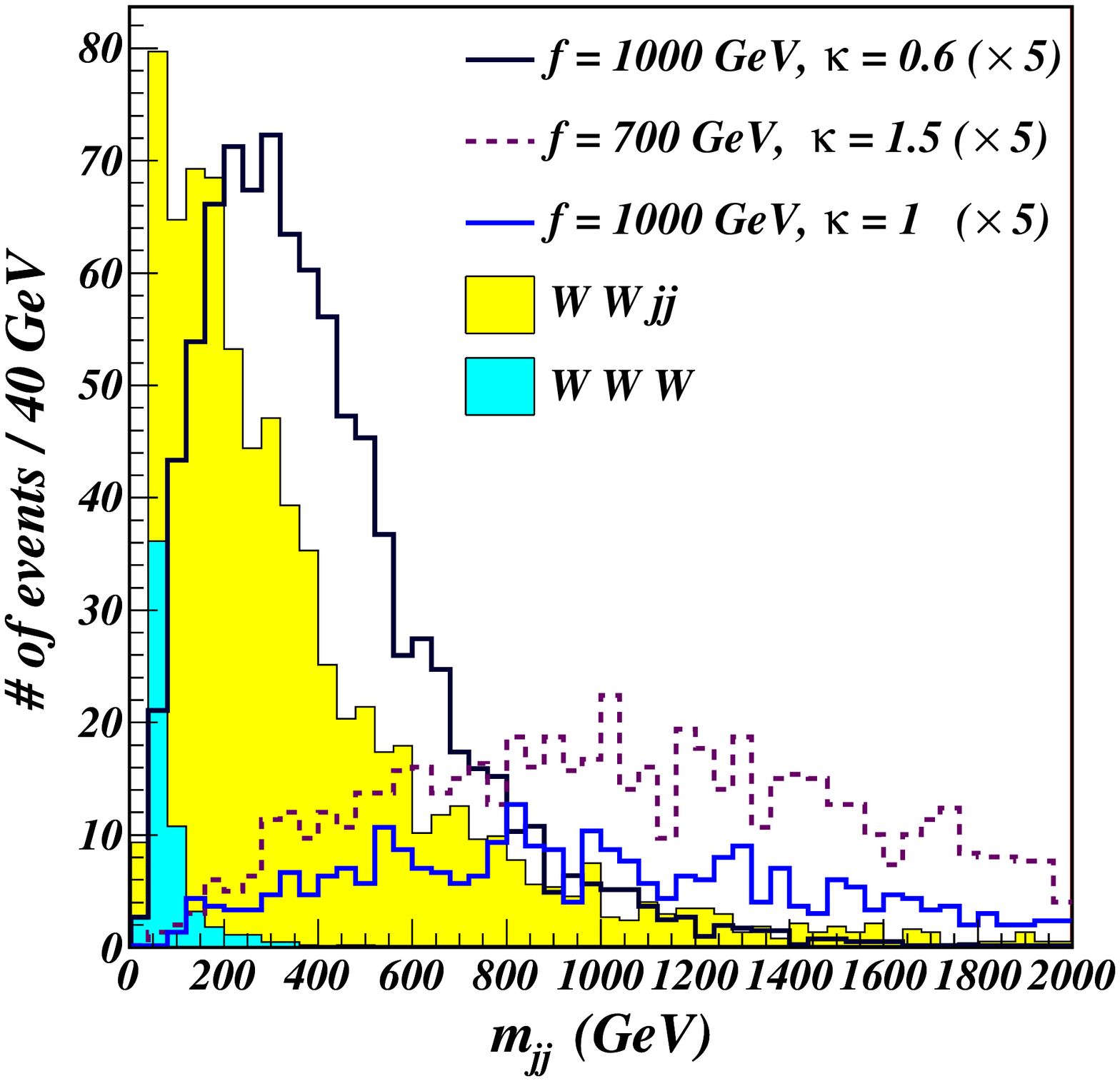,width=.5\textwidth} \hspace{-.5cm}
\epsfig{file=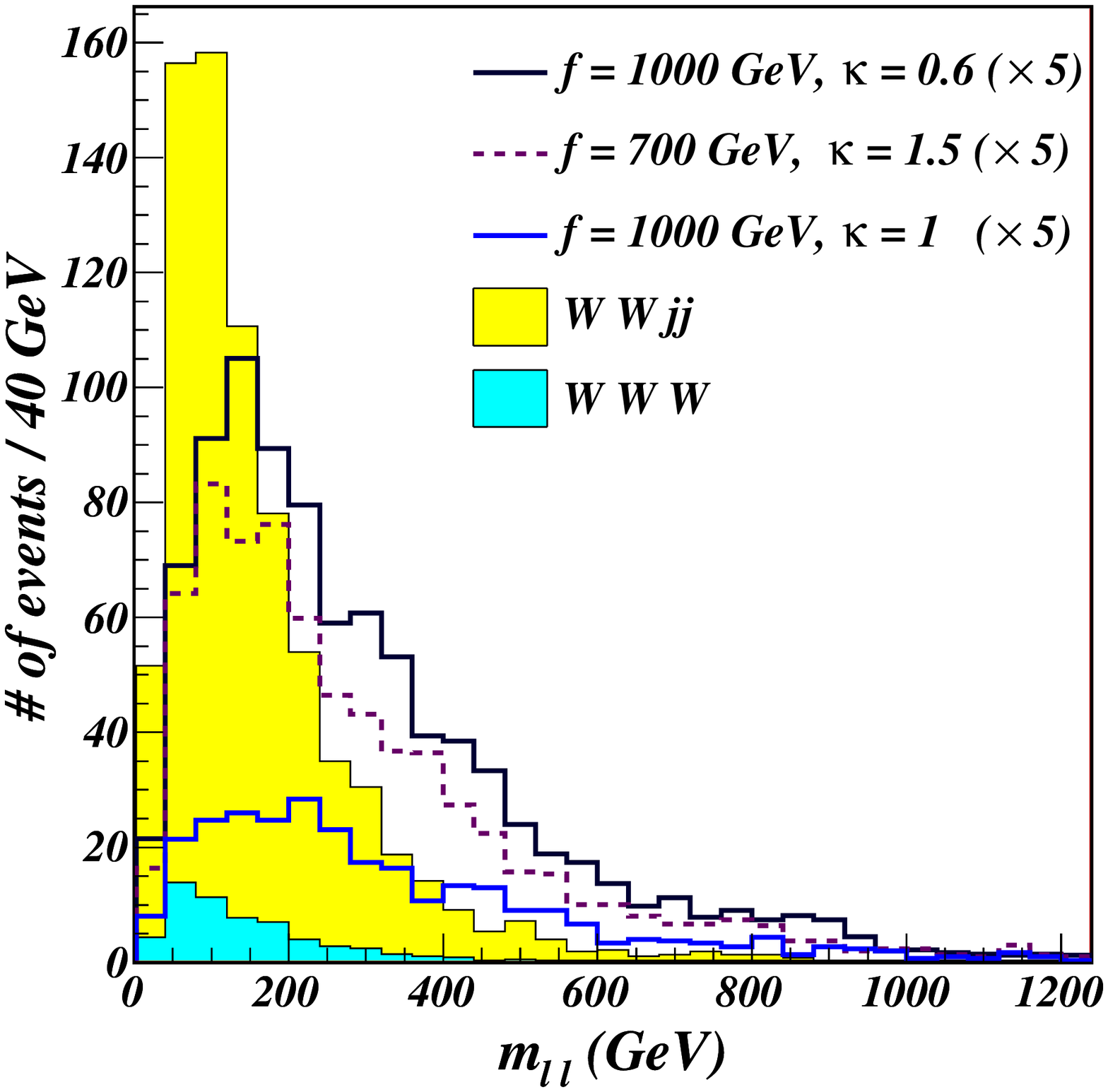,width=.5\textwidth} \\
\caption{\sl $\et$ distribution (left panel) and same-sign dilepton
  invariant mass ($m_{\ell^\pm \ell^pm}$) distribution
  (right panel) for signal ($\times 5$) and SM background in the
  channel $p p \to \ell^\pm \ell^\pm j j \et$.  In plotting these
  distributions we have assumed the LHC luminosity to be ${\cal L} = 100$
  fb$^{-1}$. }
\label{fig:todd:4}
\ecen
\end{figure}
%%=========================================%%

%%=========================================%%
%% Figure 
\begin{figure}[htb]
\bcen
\epsfig{file=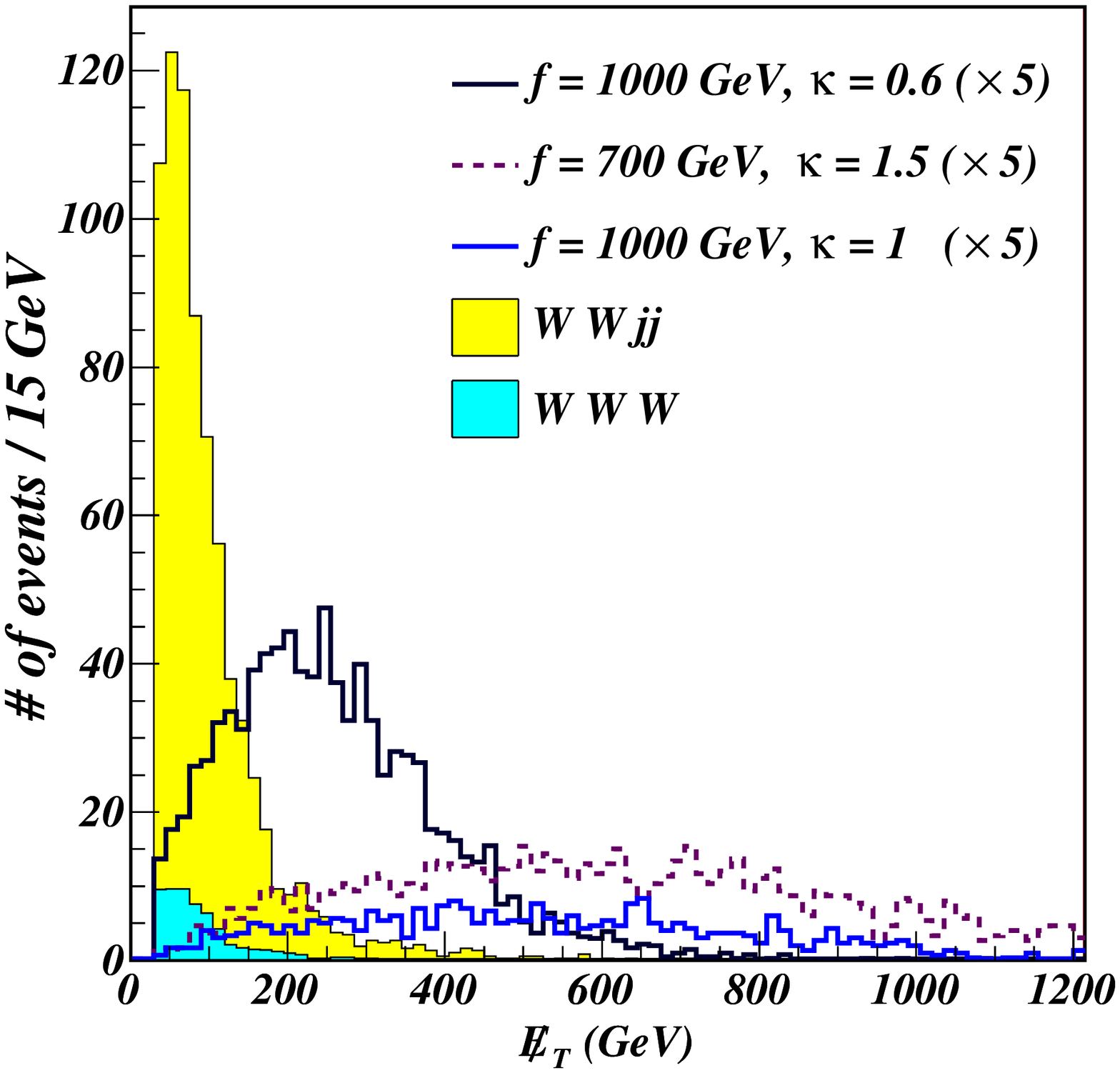,width=.5\textwidth} \hspace{-.5cm}
\epsfig{file=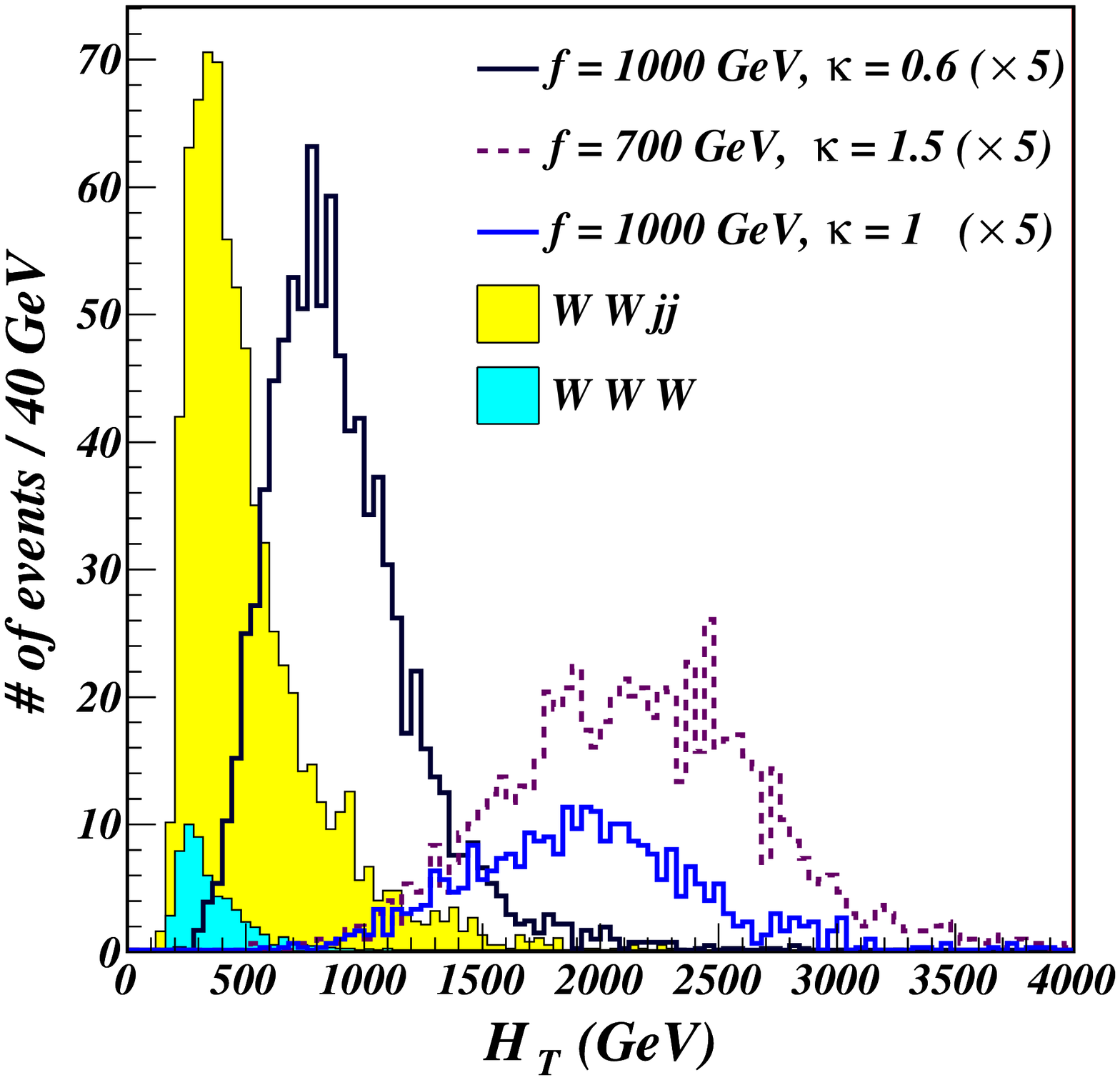,width=.5\textwidth} \\
\caption{\sl$\et$ and  $H_T$ distribution
  (right panel) for signal ($\times 5$) and SM background in the
  channel $p p \to \ell^\pm \ell^\pm j j \et$ for an LHC luminosity of
  ${\cal L} = 100$ fb$^{-1}$. }
\label{fig:todd:6}
\ecen
\end{figure}
%%=========================================%%

This signature is generated by the production mechanism given in
Eq.\ (\ref{eq:prod:3}) and the decay chains given in
Eqs.\ (\ref{decay-chain}). The same sign dilepton signature has
relatively less SM backgrounds and hence could be more useful to
search for the $T$-odd quarks in pair production at LHC. 

\noindent The possible backgrounds to our signal are:
\begin{itemize}
\item{} $p p \to W^\pm W^\mp W^\pm $, where one of the $W$-bosons decays
  into jets and the two same sign $W$-bosons decays leptonically. The
  production cross section for this process is 127 fb. 
\item{} $p p \to W^\pm W^\pm j j$ with both W-bosons decaying
  leptonically. The production cross section for this process is $\sim
  420$ fb. 
\end{itemize}

\noindent The {\sl pre-selection} cuts used are: 
\begin{itemize}
\item[(a)] exactly two light jets with $p_T^j > $ 30 GeV and $|\eta| <
  2.5$;
\item[(b)] exactly two leptons of same sign with $p_T^\ell > 15$ GeV and
  $|\eta| < 2.5$; 
\item[(c)] a minimum missing energy threshold of $\et > 30$ GeV. 
\end{itemize}

In Fig.\ \ref{fig:todd:4}, we have shown the dijet invariant mass
($m_{jj}$) for light jets and the same-sign dilepton invariant mass 
($m_{\ell^\pm \ell^\pm}$) distributions. 
 We impose a cut on jet invariant mass  of
 \beq
|m_{jj} - M_W| > 20 GeV\ .
\eeq
This cut helps reducing the background coming from $W^\pm W^\mp
W^\pm$. In Tab.\ \ref{table:tq:2}, we have shown the effects of the
pre-selection cuts and the cut on the dijet invariant mass on the signal
and backgrounds. 

% In Fig.\ \ref{fig:todd:6} we have shown the $\et$ distribution. 
As previously
argued, in LHT models we expect to have hard $\et$, and
hence one can substantially reduce the SM backgrounds by using a hard
$\et$ cut. This is also evident from the $\et$ distribution shown in
Fig.\ \ref{fig:todd:6}. We have accordingly shown the 
number of expected events for $\et > 200, 300, 400$ GeV.

%%+++++++++++++++++++++++++++++++++++++++++++++++++++++++++++++++
%% Table 
\begin{table}[htb]
\bcen
\caption{Same-sign dilepton results in the channel $p p \to \ell^\pm
  \ell^\pm j j$ for ${\cal L} = 100$ fb$^{-1}$. }
\label{table:tq:2}
\begin{tabular}{| c | c | c | c || c | c |} \hline 
Parameter set $\Rightarrow$ & $f = 1000$ &  $f = 700$  & $f
= 1000$ & SM & SM \\ 
Cuts $\Downarrow$  & $\kappa = 0.6$ & $\kappa = 1.5$ & $\kappa = 1$ 
& $W^\pm W^\pm jj$  & $W^\pm W^\pm W^\mp$    \\ \hline 
Production $\sigma$ (fb)    & 235.1 & 240.7 & 80.1  &    &
\\
Pre-selection               & 180.5 & 140.9 & 58.5  & 747.2 & 59.1  \\ 
$|m_{jj} - M_W| > 20$ GeV    & 173.9 & 140.6 & 58.5  & 651.2 & 20.3  \\
\hline 
$\et > 100$                 & 155.7 & 138.5 & 56.6  & 236.1 & 5.8   \\
${\cal S}$                  &   9.1 &  8.2  &  3.5  &       &       \\ \hline
$\et > 200$                 & 108.4 & 129.4 & 51.6  & 57.8  & 0.9    \\
$\et > 300$                 & 57.7  & 117.4 & 45.1  & 22.2  & 0.3    \\
$\et > 400$                 & 24.9  & 103   & 37.5  &  9.6  & 0.1   \\
${\cal S}$                  & 6.2   & 18.6  &  8.6  &       &       \\ \hline
\end{tabular}
\ecen
\end{table}
%%+++++++++++++++++++++++++++++++++++++++++++++++++++++++++++++++

Finally, for the signature  $\signaturetwo$, we have shown the total
transverse energy distribution ($H_T$). As previously argued, this
distribution could be useful in having a estimate of the $T$-odd 
quark masses. 

%%%+++++++++++++++++++++++++++++++++++++++++++++++++++++++++++++++%%%
%%% Section : Conclusions                                         %%%

\section{Conclusions \label{section:5}}

In this paper, we have analyzed the signatures of the pair production
of the first two generations of $T$-odd quarks in the context of the LHT
model. We first showed that for a reasonable range of input parameters,
these quarks can be produced in large numbers both at the low energy
(10 TeV) run and the full 14 TeV run of the LHC. The $T$-odd quark
masses depend only on $\kappa$ and $f$.
The branching fractions for
the decays of these quarks crucially depend on $\kappa$: 
for large values ($\kappa > 0.462$), the main mode is in
 $W_H$ plus a light quark and it can contain a lepton coming 
 from the subsequent leptonic decay of the $W$.

In this work, we considered both same-sign and opposite-sign dilepton
signatures: (a) opposite-sign dilepton pairs, $p p \to (\bar{Q}_H Q_H,
Q_H Q'_H) \to (q \bar{q}, q q') W_H^\pm W_H^\mp \to j j \ell^\pm
\ell^\mp \et$,  and (b) same-sign dilepton pairs, $p p \to Q_H Q'_H \to q
q' W_H^\pm W_H^\pm \to j j \ell^\pm \ell^\pm \et$. We used CalcHEP to
generate the signal and further interfaced it with PYTHIA.
For realistic estimates of signal and backgrounds, we used the
fast ATLAS detector simulator ATLFAST. We also gave the possible  
$K$-factors for signal and background processes. As could be seen, the
relevant $K$-factors can give substantial enhancements in the production
rates of the first two generations of $T$-odd quarks at LHC. 

To quantify our results, we have also shown the significance of the
results for three sets of LHT model input points.  
As signal and background events after the cuts are smaller in number,
we have to use the Poisson statistics to estimate the significance of
the results. 
To quantify our results for the set of input parameters chosen, we use
a significance estimator \cite{Ball:2007zza}
$$
{\cal S} = \sqrt{2 \left\{ n_0 \ ln\left(1 + \frac{s}{b}\right) - s \right\}},
$$
where $b$ is the expected number of background events and $n_0$ is the
number of observed events. Accordingly, the signal is defined as $s =
n_0 - b$. This estimator is based on a log-likelihood ratio and follows
very closely the Poisson significance.
We used the minimum set
of pre-selection cuts as defined in Sec.\ \ref{section:4} for
signal and background processes. We tried to analyze various
distributions that could help in selecting secondary set of cuts to
improve signal to background rates. We found that the hard cuts on 
$\et$ could be useful in extracting the signal from the
backgrounds. Accordingly, we showed the results of signal and
background events for a $\et$ cut of 100, 200, 300 and 400 GeV.
The summary of the number of events after imposing various selection
cuts were given in Tab.\ \ref{table:tq:1} (opposite-sign dilepton) and
Tab.\ \ref{table:tq:2} (same-sign dileptons). 
As can be seen from the summary tables of the results, the
signal to background ratio can be significantly improved in the case of
same-sign dileptons and even further by using the
$H_T$ distribution. In case of opposite-sign
dileptons, the $t \bar{t}$ backgrounds make the $H_T$-distribution peak
at much larger values, and hence this distribution could be useful to
improve signal to background, if the masses of $T$-odd quarks are much
higher as compared to the SM top-quark. In the case of same-sign leptons, the
backgrounds come from gauge bosons, and hence the $H_T$-distribution peaks
at relatively lower values. Hence this distribution is much more
useful to suppress backgrounds. At this point we would like to note,
as is evident from Fig.\ \ref{fig:br1}, that the second leading decay channel
of $T$-odd quarks is $Q_H \to Z_H q$. Although this is not the
dominant decay chain, it can lead to some very interesting
signatures \cite{Choudhury:2006mp}. This issue will be addressed in a
future work \cite{progress}.  
 In Tabs.\ \ref{table:tq:1}
and \ref{table:tq:2}, we have also given the expected significance
(${\cal S}$) of the results for three sets of LHT input parameters. The
results indicate that the significance of the resuls in the signal (b),
same-sign dileptons, provides a very interesting discovery potential, as
the backgrounds in this channel are very low. This channel can also
give a good estimate of the masses of $T$-odd quarks from the
$H_T$-distribution. As can be seen from the results, one must optimise
the secondary selection cuts to further improve the signal rates as
compared to the backgrounds. This is beyond the scope of the present
paper. To summarize, the pair production of $T$-odd quarks can probe
a substantial $(f, \kappa)$ region of the LHT model parameter
space. For some reasonable values of $\kappa$ using the same-sign
dilepton channel described above, one can probe the symmetry breaking
scale of the LHT model ($f$) up to the TeV range. 
We hope that this study will motivate the LHC collaborations to
search for the first two generations of $T$-odd quarks at LHC.   

%%%+++++++++++++++++++++++++++++++++++++++++++++++++++++++++++++++%%%
%%% Section : Acknowledgements                                     %%%

\section*{Acknowledgements} 
We would like to thank Satyaki Bhattacharya, John Campbell, Debajyoti Choudhury,
and Peter Skands for useful discussions/communications. We also like
to thank A. Belyaev for providing us the original LANHEP files
for the LHT model \cite{Belyaev:2006jh}.  
AD and MK are supported by the ANR project ANR-06-JCJC-0038-01 and the
Theory-LHC-France initiative of the CNRS/IN2P3.
The work of SRC was supported by Ramanna fellowship of Department of
Science \& Technology (DST), India.  

%%%%%%%%%%%%%%%%%%%%%%%%%%%%%%%%%%%%%%%%%%%%%%%%%%
%% References 

\end{document}